# Chirality control by electric field in periodically poled MgO-doped lithium niobate

Lei Shi, Linghao Tian, Xianfeng Chen*

Department of Physics, The State Key Laboratory on Fiber Optic Local Area Communication Networks and Advanced Optical Communication Systems, Shanghai Jiao Tong University, 800 Dongchuan Rd., Shanghai 200240, People's Republic of China

*xfchen@sjtu.edu.cn

We study the chirality of periodically poled MgO-doped lithium niobate (MgO:PPLN) by electro-optic (EO) effect. It shows that optical propagation is reciprocal in MgO:PPLN when quasi-phase-matching (QPM) condition is satisfied, which is similar to natural optical active material like quartz. The specific rotation of MgO:PPLN by EO effect is shown to be proportional to the transverse electric field, making large polarization rotation in optical active material with small size possible. We also demonstrate that the chirality of MgO:PPLN can be controlled by external electric field.

Optical activity is the turning of the polarization plane of linearly polarized light about the direction of motion as light travels through certain materials. It occurs in solutions of chiral molecules such as sucrose, spin-polarized gases of atoms or molecules, and solids with rotated crystal planes such as quartz. It is widely used in the sugar industry to measure syrup concentration, in optics to manipulate polarization, in chemistry to characterize substances in solution, and in optical mineralogy to help identify certain minerals in thin sections. In an optically active material, optical propagation is reciprocal and the polarization direction rotates in the same sense (e.g. like a right-handed screw) during the forward and backward pass. The rotation angle of polarization direction in an optical active material as quartz is $\beta = \alpha L$, where α is the specific rotation, and L is the path length of light in the material. The specific rotation of a pure material is an intrinsic property of that material at a given wavelength and temperature. A positive value corresponds to dextrorotatory rotation and a negative value corresponds to levorotatory rotation.

The periodically poled lithium niobate (PPLN), an artificial nonlinear material, is receiving more and more attention owing to its outstanding nonlinear optical properties. The even order nonlinear coefficients, such as electro-optics coefficient and photo voltaic coefficient, are periodically modulated due to the ferroelectric domain inversion. It has been widely used in frequency conversion, pulse shaping, optical switching, and other nonlinear optical processes by QPM technique[1-6]. Recent research has shown that the polarization direction of an incident linear polarized light, passing through a PPLN with transverse EO effect, rotates with the increment of the external electric field when the operating wavelength satisfies the QPM condition. EO effect of PPLN has been applied to various fields like EO Solc-type wavelength filter[7, 8], logic gates[9], optical isolator[10], high frequency modulators[11], scanners[12], lens[13] and polarization state generator[14] since its discovery. However, the chirality of this kind of optical rotation is still unclear and worth studying for its splendid further applications in optical signal control.

In this letter, we give a deep insight into the chirality of the PPLN with EO effect. It shows that optical propagation is reciprocal in PPLN, similar to the general optical active medium as quartz, in which the polarization direction twists in the same sense during the forward and backward pass. The specific rotation of PPLN by EO effect is shown to be proportional to the transverse electric field, making it more convenient to be flexibly adjusted based on practical demand. We also demonstrate that the chirality of PPLN can be controlled by the external electrical field.

EO effect of PPLN has its origin in the work by Lu et al[15]. Their study shows that when a transverse electric field is applied along the +Y axis of PPLN, as shown in Fig.1, the refractive-index ellipsoid deforms. Consequently, the Y and Z axis of the refractive-index ellipsoid rotate an angle of $\theta$ around X axis. The rocking angle $\theta$ is proportional to the external electric field and is given by $\theta \approx \gamma_{51} E_y / [(1/n_e)^2 - (1/n_o)^2]$, where $n_o$ and $n_e$ are refractive indices of the ordinary and the extraordinary wave, respectively; $E_y$ is the external electric field, and $\gamma_{51}$ is the EO coefficient. Since all elements of the EO tensor have different signs in different domains, the azimuth angle of the new optical axes rock from

$+\theta$ or $-\theta$ successively when an external electric field is applied. When the QPM condition is satisfied, each domain serves as a half-wave plate with respect to the input light. After passing through a stack of rotated half-wave plates, the optical polarization plane of the input light will rotate continually and emerge finally at an angle of 2Nθ, where N is the number of domains. Furthermore, with the increment of the electric field, the rotation angle of the input light will rotate correspondingly, since the optical axes of the half-wave plates rotate continually with the electric field, which is similar to a birefringence half-wave plate rotated manually.

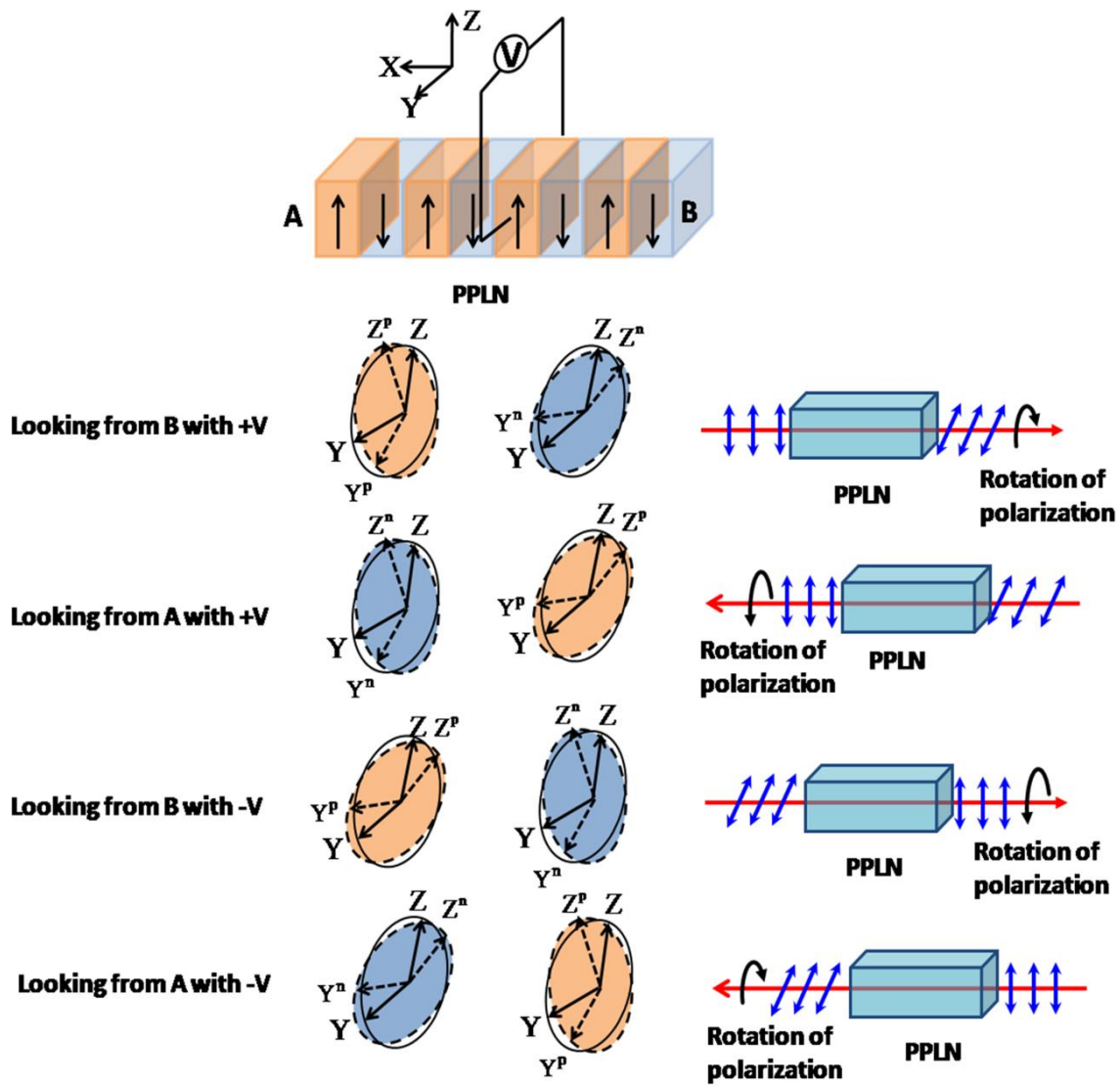

FIG.1 (Color online) Schematic diagram of PPLN crystal, the rotation of principal axes and final rotation direction of the polarization plane when a transverse electric field is applied. The arrows inside the PPLN indicate the spontaneous polarization directions. Deformation of the index ellipsoid is observed from port A and B (representing light travelling along +X and –X axis) under a $\pm V$

voltage (±represent transverse electric field along ±Y axis). X,Y and Z represent the principal axes of the original index ellipsoid, and $Y_{p,n}$, $Z_{p,n}$ are the perturbed principal axes of the positive and negative domain, respectively.

When an external voltage V is applied along +Y axis of the PPLN, as presented in Fig.1, looking along +X axis, the Y and Z axis of the index ellipsoid rotate an angle of θ left-handedly and right-handedly in the positive and negative domains, respectively. When QPM condition is satisfied, the polarization direction of an incident linearly polarized light travelling along –X axis will rotate an angle of 2Nθ (N is the number of domains) right-handedly at the output side. However, looking along the -X axis, we see that the Y and Z axis of the index ellipsoid rotate an angle of θ right-handedly and left-handedly in the positive and negative domains, respectively. And the overall effect is that a linearly polarized light travelling along +X also rotates an angle of 2Nθ right-handedly at the output side. Thus, the polarization direction twists in the same sense during the forward and backward pass, and the optical propagation is reciprocal in PPLN, which is similar to optical active material as quartz. When the external voltage V is applied along -Y axis of the PPLN, the rotation directions of the Y and Z axis of the index ellipsoid also reverse, and the polarization direction rotates left-handedly during the forward and backward pass. The chirality of PPLN is thus controlled by the external electrical field.

The rotation angle of the polarized light after passing through the PPLN is $\beta = 2\frac{L}{\Lambda}\frac{\gamma_{51}E}{(1/n_e)^2 - (1/n_o)^2} = \alpha L$, where $\Lambda$ and L are respectively domain thickness and length of the PPLN. The specific rotation, defined as $\alpha = \frac{2}{\Lambda}\frac{\gamma_{51}E}{(1/n_e)^2 - (1/n_o)^2}$, is relevant to the wavelength, temperature and material, and is electric field adjustable. It is thus advantageous over optical active material in that the specific rotation can be adjusted based on practical demand. Large optical rotation in material with small size is then at hand. For a MgO:PPLN with 3582 domains and domain period of $20.1\,\mu m$, we measured the specific rotation under different electric field with the working wavelength of 1568.5nm at $22^0 C$. The result is shown in Fig.2. The specific rotation increases linearly with the electric field, and reaches $0.87^0$/mm under an external electric field of 3kV/cm. The MgO:PPLN used in our experiment is one with dimensions of 36mm × 9.2mm × 0.5mm. It is comprised of 10 gratings, with periods ranging from 19.5μm to 21.3μm. When the transverse electric voltage is applied on the MgO:PPLN, much larger electric voltage is actually required to guarantee the expected rotation angle in domains of target period, resulting in a relatively smaller specific rotation under a defined electric voltage. In the waveguide configuration, the width of the PPLN can be as small as 10μm. The specific rotation can be as large as $2.43^0$/mm under an electric voltage of 1V, which is very attractive.

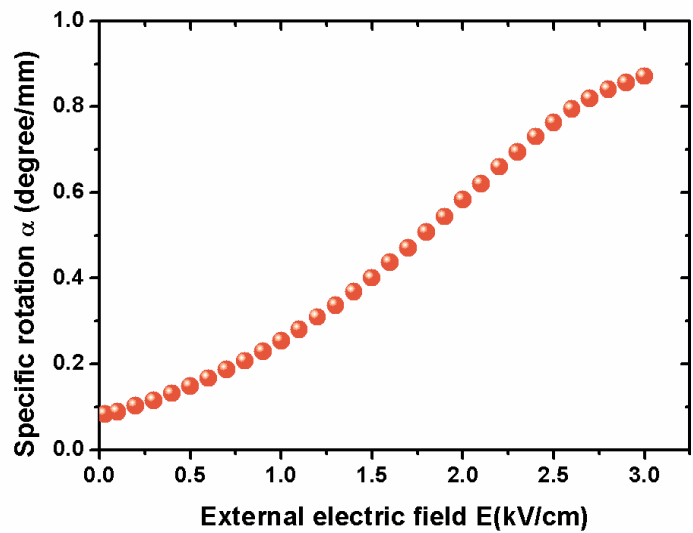

FIG.2. Experimental measurement of the specific rotation versus electric field. The input wavelength is 1568.5nm at $22^0$C, and the MgO:OPPLN is one with the period of 20.1μm and 3582 domains.

We designed an experiment to demonstrate the chirality of PPLN, as shown in Fig.3. A tunable laser worked as light source. Two polarization-beam-spliters (PBSs) were set perpendicularly to work as polarizer and analyzer. A MgO:PPLN crystal and a dextrorotatory $45^0$ quartz were placed between the two PBSs. The MgO:PPLN and working environment are the same as that in the specific rotation measurement. We measured the output power of light travelling along +X and -X axis with an external electric field applied along +Y or −Y axis, which is expressed as $T = \sin^2(\pi/4 \pm 2N\theta)$[7]. The results are presented in Fig.4. The pink and blue curve represent transmissions of light travelling along +X (forward wave) and −X direction (backward wave), respectively. Apply the electric field along +Y axis, we get the transmissions shown in Fig.3 (a).With the help of the dextrorotatory quartz, transmission curves of light present a form of sinusoidal function, getting the maxima first with the increment of the electric field, which means a rotation of the polarization direction of about $45^0$ right-handedly from the MgO:PPLN. While when an electric field along –Y axis is employed, the transmission curves appear a cosine-function shape. They reach the minima first when the electric field increases from 0, as shown in Fig.4(b), indicating that light undergoes a left-handed rotation from the MgO:PPLN. From Fig.4(a) and (b), we get the conclusion that the change of chirality of MgO:PPLN can be achieved by altering the direction of the applied electric field.

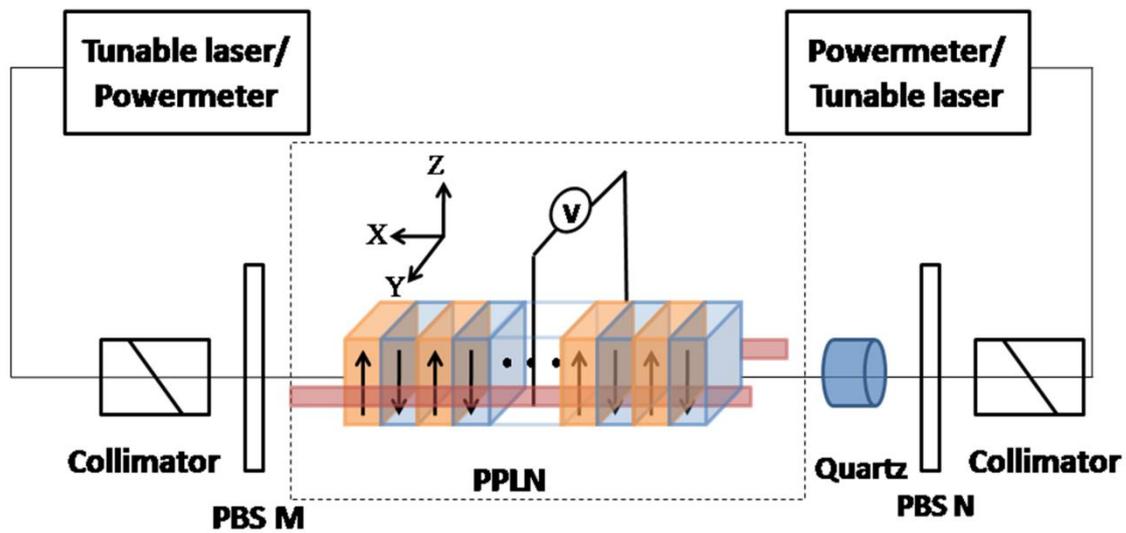

FIG.3. (Color online) Experimental setup for studying the optical activity of MgO:PPLN. The MgO:PPLN crystal is Z cut. Two PBSs are perpendicularly placed: M Z-oriented, and N Y-oriented. A uniform electric field is applied along the +Y or –Y axis.

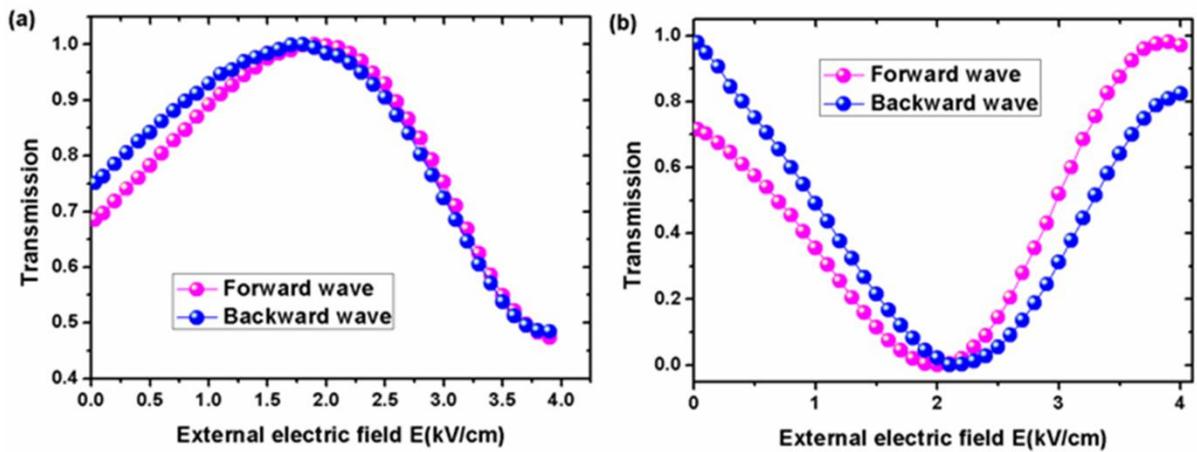

FIG.4. (Color online) Normalized transmission of light traveling in +X (forward wave) and –X direction (backward wave) in MgO:PPLN. (a) The transverse electric field is applied along +Y axis.

Compare the pink and blue curve in Fig.4(a) or (b), we find that they are almost of the same shape, revealing that light experience the same rotation process, and verifying that optical propagation is reciprocal in MgO:PPLN. The small shift between these two curves is probably caused by the temperature fluctuation during the experiment, which induces the variation of the QPM wavelength. The electric field where the maximum and minimum transmissions occur is actually larger than the theoretical expectation, and is due to several reasons. The overriding reason is that the external electric field is generated by use of a pair of parallel copperplates, which requires extreme closeness to the MgO:PPLN crystal. During the experiment we found that by slightly pressing the copperplates toward the MgO:PPLN crystal, less external electric field was required to obtain the same rotation angle. However, we did not perform strong pressure on them and the pressure was different for each time, which resulted in incomplete closeness and consequently led to the discrepancy. Secondly, the $45^0$ dextrorotatory quartz we used is fabricated at the wavelength of 1550nm. When the wavelength is 1568.5nm, the rotation angle is $43.88^0$, which can be obtained from the formula $\varphi = \alpha * d$, where $\varphi$ is the rotation angle, $\alpha$ is the specific rotation, and d is the thickness of quartz. Thereupon, the shift of rotation angle from the quartz gives rise to a shift in the applied electric field from the theoretical anticipations. The temperature fluctuation mentioned above is also responsible for the shift.

In conclusion, we analyzed the chirality of MgO:PPLN with EO effect. With a transverse electric field applied on the MgO:PPLN, the optical propagation in MgO:PPLN is reciprocal, which is similar to natural optical activity. The specific rotation of MgO:PPLN by EO effect is shown to be proportional to the transverse electric field, that the rotation angle can be adjusted as request. Besides, we also demonstrate that the chirality of MgO:PPLN can be controlled by external electrical field.

This research was supported by the National Natural Science Foundation of China (Grant No. 61125503, 61078009), the National Basic Research Program "973" of China (Grant No. 2011CB808101), the Foundation for Development of Science and Technology of

Shanghai (Grant No. 11XD1402600) and the Open Fund of the State Key Laboratory of High Field Laser Physics.